\documentclass[preprintnumbers,amsmath,amssymb,showpacs,twocolumn]{revtex4}

\usepackage{graphicx}
\usepackage{dcolumn}
\usepackage{bm}

\begin{document}

\title{ Conclusive quantum state classification }

\author{Mei-Yu Wang$^{1}$, Feng-Li Yan$^{1,2}$}

\affiliation {$^1$ College of Physics  and Information
Engineering, Hebei Normal University, Shijiazhuang 050016,
China\\
$^2$ CCAST (World Laboratory), P.O. Box 8730, Beijing 100080, China\\}

\date{\today}

\begin{abstract}
  In this paper, we have considered the problem of general conclusive quantum state classification;
  the necessary and sufficient conditions for the existence of conclusive classification strategies have also
  been presented. Moreover, we have given the upper bound for the maximal success probability.
\end{abstract}


\maketitle

\section{Introduction}

In classical mechanics, it is possible in principle to determine the state of any physical system. The state of
a classical system is described by its canonical coordinates and momenta, which can be measured simultaneously
to arbitrary precision. In  quantum mechanics, however, the state of a system is represented as a vector in a
Hilbert space, and is not  itself an observable quantity. Precise determination of a completely unknown state
vector is precluded by the nature of the quantum measurement process. It is  only when the state belongs to a
known orthogonal set that it can be infallibly determined by a standard Von Neumann measurement.

When confronted with the problem of trying to discriminate among non-orthogonal states, we must accept that no
strategy will correctly reveal the state of the system with unit probability. Two types of discrimination are
usually considered for non-orthogonal states in the literature: one is conclusive discrimination (also called
unambiguous discrimination) and the other is inconclusive discrimination. The main difference between them is
that the former allows the "don't know" claim but no wrong answers while the latter permits incorrect judgement
of the system. Conclusive discrimination originated from the works of Ivanovic \cite {s1}, Dieks \cite {s2} and
Peres \cite {s3}, who initially distinguished two non-orthogonal states $|\phi\rangle$ and $|\psi\rangle$ with
the same prior probabilities and derived the maximum probability of success called IDP limit as
$1-|\langle\phi|\psi\rangle|$. Subsequently, Jaeger and Shimony \cite {s4} generalized it to the case of unequal
a prior probability and obtained the results as $1-2\sqrt {pq}|\langle\phi|\psi\rangle|$ or
$p(1-|\langle\phi|\psi\rangle|)$ in case $p\geq q$. In reality, various measurement schemes have been suggested,
and some bounds on the probability for conclusive discrimination have derived. Peres and Terno \cite {s5}
discussed in detail the problem of optimal distinction of three states having arbitrary a prior probabilities.
 Chefles \cite {s6} showed that a set of quantum states is amenable to conclusive discrimination, if and only if
 they are linearly independent. The optimal conclusive discrimination among linearly independent symmetric
 states was investigated by Chefles and Barnett \cite {s7}.
    But indeed the optimal solution is unknown yet. However, deriving some upper bounds on the success
    probability for conclusive discrimination among states is possible \cite {s8,s9}. Also, it is worth
    mentioning that many of the theoretically discovered optimal schemes for discrimination have been
    experimentally realized \cite {s10,s11,s12}.

    Quantum state discrimination is to reveal the full information of a system, but sometimes we only require
    part of information. So we turn our attention to the following variant of problem. Instead of discriminating
    among all states, we ask what happen if we just want to discriminate among subsets of them. In this class of
    problems, we know that a given system is prepared in one of $N$ quantum states, which constitute a state set
    $S$, but we do not know which one it is. What we want to do is to determine which subset of $S$ the state
    belongs to. Here, the subsets of $S$ are denoted by $S_1,S_2,\cdots,S_n$ respectively, which satisfy
    $\cup_{i=1}^n S_i=S$ and $S_i\cap S_j={\o}$ for $i\neq j$. In fact, a subset may be viewed as a class of quantum
    states, so we call this process  the quantum state classification. When the subsets are not mutually
    orthogonal, the classification can not be done with unit probability of success. Like the state discrimination,
     there are two types of quantum state classification: conclusive classification \cite {s13} and inconclusive
     classification \cite {s14}. In [13], the authors considered the case of conclusive discrimination between
     $\{|\psi_1\rangle\}$ and $\{|\psi_2\rangle, |\psi_3\rangle\}$ with respective a prior probabilities
     $\eta_1,\eta_2,\eta_3$. The optimum success probability is derived. Bergou et al \cite {s15,s16}
     generalized it to the case between $\{|\psi_1\rangle\}$ and $\{|\psi_2\rangle, |\psi_3\rangle, \cdots,
     |\psi_N\rangle\}$.

     In this paper, we will consider the problem of general conclusive quantum state classification, the
     necessary and sufficient conditions for the existence of conclusive classification strategies is presented.
     Moreover, we give the upper bound for the maximal success probability of conclusive state classification
     among $n$ subsets.

\section{Conditions of conclusive state classification}

Suppose we are given a quantum system prepared in the state
$|\psi\rangle$, which is guaranteed to be a member of a set $S$ of
$N$ known non-orthogonal states. The state set $S$ can be divided
into $n$ subsets: $S_1, S_2, \cdots, S_n$. We have
\begin{equation}
\bigcup_{i=1}^nS_i=S,   S_i\cap S_j={\o}. (i\neq j).
\end{equation}
Without loss of generality, we denote by $|\psi_{ik}\rangle$ the $k$th element in $S_i$ $(i=1,2,\cdots, n;
k=1,2,\cdots, m_i.)$ with a prior probability $\eta_{ik}$. Before finding a strategy for conclusive quantum
state classification, a natural query raised is when the state $|\psi_{ik}\rangle$ can be conclusive classified
with a non-zero success probability. We have the following theorem.

{\it Theorem}: The necessary and sufficient conditions for the existence of conclusive quantum state
classification is that there exists at least a state $|\psi_{i_0k_0}\rangle$ can not be linearly represented by
those states in other subsets.

{\it Proof}: If the quantum states can be classified conclusively,
there must be a generalized measurement (also called POVM) that is
described by  a set of linear operators $\{A_m\}$: $n$ operator
$A_m$ correspond to the subsets $S_1, S_2, \cdots, S_n$
respectively, and one operator $A_I$ generates inconclusive
results. These operators must satisfy the completeness equation:
\begin{equation}
A_I^+A_I+\sum_mA^+_mA_m=I.
\end{equation}
Here $I$ is the identity operator. The {\it m}th outcome should
only arise if the initial state is in $S_m$, implying the
following constraint:
\begin{equation}
\langle\psi_{ik} |A^+_mA_m|\psi_{ik}\rangle=P_{ik}\delta_{mi}.
\end{equation}
In Eq. (3), $P_{ik}$ is the probability of successfully assigning $|\psi_{ik}\rangle$ to the subset $S_i$, and
$\langle\psi_{ik} |A^+_IA_I|\psi_{ik}\rangle$ is the probability of failing to assign $|\psi_{ik}\rangle$. So
the average probability of success is
\begin{equation}
P=\sum_{i=1}^n\sum_{k=1}^{m_i}\eta_{ik}\langle\psi_{ik}
|A^+_iA_i|\psi_{ik}\rangle.
\end{equation}
If $P\neq 0$, there exists at least one nonzero term in the summation of Eq.(4). Without loss of generality, we
suppose $\langle\psi_{i_0k_0} |A^+_iA_i|\psi_{i_0k_0}\rangle\neq 0.$ We claim $|\psi_{i_0k_0}\rangle$ can not be
linearly represented by the states in $S-S_{i_0}$. Indeed, suppose
\begin{equation}
|\psi_{i_0k_0}\rangle=\sum_{i\neq i_0}\sum_{k=1}^{m_i}C_{ik}|\psi_{ik}\rangle,
\end{equation}
 then, from Eq.(3) we have $\langle\psi_{i_0k_0} |A^+_{i_0}A_{i_0}|\psi_{i_0k_0}\rangle=\sum_{i,i'\neq
 i_0}\sum_{k,k'}C_{ik}^*C_{i'k'}\langle\psi_{ik}|A^+_{i_0}A_{i_0}|\psi_{i'k'}\rangle=0$, which contradicts $\langle\psi_{i_0k_0}
 |A^+_{i_0}A_{i_0}|\psi_{i_0k_0}\rangle\neq 0$. Therefore, the above claim is correct.

 On the other hand, if there exists a quantum state $|\psi_{i_0k_0}\rangle$ can not be linearly represented by
 the states in $S-S_{i_0}$, we can rewrite it as
\begin{equation}
|\psi_{i_0k_0}\rangle=\sum_{i\neq
i_0}\sum_{k=1}^{m_i}C_{ik}|\psi_{ik}\rangle+d|\psi_{i_0k_0}^\perp\rangle,
\end{equation}
where $|\psi_{i_0k_0}^\perp\rangle$ is orthogonal to each $|\psi_{ik}\rangle$ $(i\neq i_0)$. Apparently $d\neq
0$.

What we need to do is to construct a set of measurement operators, which satisfy Eq. (2)-(3), and $P\neq 0$. Let
\begin{eqnarray}
&&A_{i_0}=|\psi_{i_0k_0}^\perp\rangle\langle\psi_{i_0k_0}^\perp|,\nonumber\\
&&A_i=0,        (i\neq i_0),\nonumber\\
&&A_I=\sqrt {I-A^+_{i_0}A_{i_0}}.
\end{eqnarray}
It is easy to verify that the above operators satisfy Eq.(2)-(3)
and
\begin{equation}
P=\sum_{i=1}^n\sum_{k=1}^{m_i}\eta_{ik}\langle\psi_{ik} |A^+_iA_i|\psi_{ik}\rangle=\eta_{i_0k_0}|d|^2\neq 0.
\end{equation}
 This has completed the proof.

That a quantum state can not be linearly represented by some
states implies this state does not lie in the space spanned by
those states. So we can give a vivid picture of the above theorem.
That is, there  exists the conclusive classification if and only
if at least one state does not be "submerged" by states in other
"classes". When every state in $S$ can be represented by quantum
states in other subsets of $S$, one can not find any measurement
strategy to complete the conclusive classification. This prevents
any information leakage to eavesdropper when he uses  measurement
strategy in the BB84 protocol of quantum key distribution between
two partners, Alice and Bob \cite {s19}. The BB84 protocol is
characterized by the fact that two complementary bases are used to
encode the bits. Alice prepares a series of qubits in a randomly
chosen eigenstates of $\sigma_z$ and $\sigma_x$. She encodes
classical bit 0 into the quantum states $|0\rangle$ or $\frac
{1}{\sqrt 2}(|0\rangle+|1\rangle)$, and encodes 1 into the quantum
states $|1\rangle$ or $\frac {1}{\sqrt 2}(|0\rangle-|1\rangle)$ by
prior mutual agreement of the parties. Then she sends the qubits
to Bob. In the process of the qubit transmission, the eavesdropper
captures the qubit and measures it, then sends it to Bob. If an
eavesdropper, Eve,  wants to obtain some conclusive information by
measurement, she must find a measurement strategy to ensure that
the state belongs to either $S_1=\{|0\rangle,\frac {1}{\sqrt
2}(|0\rangle+|1\rangle)\}$ or $S_2=\{|1\rangle,\frac {1}{\sqrt
2}(|0\rangle-|1\rangle)\}$. But it is easy to check that every
state of $S_1(S_2)$ can be linearly represented by the states in
$S_2(S_1)$. So no measurement strategy can conclusively determine
the state which subset it belongs to. Therefore Eve can not obtain
any conclusive information by measurement strategy in BB84
protocol.

\section{The upper bound on average success probability}

Although the measurement described by Eq.(7) can complete the task
of conclusive classification among quantum states
$|\psi_{ij}\rangle\in S$, the success probability is not optimal
in general. In the following, we will give the upper bound for the
success probability of any classification strategy.

Every classification can be described by a so-called "generalized measurement" based on POVM \cite {s17}.
 Using Neumark's theorem, a POVM can be implemented in the following way \cite {s18}. We first embed the system
 in a larger Hilbert space $\mathcal{K}$ consisting of original system space $\mathcal{H}$, and an auxiliary Hilbert space called
 the ancilla $\mathcal{A}$. We take $\mathcal{K}$ to be a tensor product $\mathcal{K}=\mathcal{H}\otimes \mathcal{A}$. Then we introduce an interaction between
 the system and ancilla corresponding to a unitary evolution on this larger space. The unitary evolution
 entangles the system degrees of freedom with those of the ancilla. Finally, a projective measurement is
 performed on the extra degrees of freedom. Due to the entanglement, a click in the ancilla detectors will also transform
 the state of the original system in a general way.

 The input state of the system is one of the $N$ quantum states, that is $|\psi_{ij}\rangle\in S$, which is now a
 vector in the subspace $\mathcal{H}$ of the total space $\mathcal{K}$, so that
 \begin{equation}
|\psi\rangle_{in}=|\psi_{ij}\rangle|P\rangle.
 \end{equation}
Here $|P\rangle$ is the initial state of the ancilla (same for all
inputs). Following the general procedure outlined in the previous
paragraph for the generalized measurement, we now apply a unitary
transformation $U$ that entangles the system with the ancilla
degrees of freedom. As a result, the input vector transforms into
the state $|\psi\rangle_{out}$. For the purpose of conclusively
classifying $N$ quantum states into $n$ subsets, we need to obtain
$n+1$ different outcomes when a projective measurement is
performed on the ancilla: $n$ outcomes tell us the input is from
which subsets, the rest one shows us the classification fails.
Thus, we require the ancilla to be $n+1$ dimensional.
\begin{equation}
U|\psi_{ik}\rangle|P\rangle=\sqrt {1-\gamma_{ik}}|\psi'_{ik}\rangle|P_i\rangle+\sqrt
{\gamma_{ik}}|\phi_{ik}\rangle|P_{n+1}\rangle,
\end{equation}
where $|P_1\rangle,|P_2\rangle,\cdots, |P_{n+1}\rangle$ are
orthogonal basis of the ancillary Hilbert space,
$|\psi'_{ik}\rangle$ is the final state of the system, and
$|\phi_{ik}\rangle$ is the failure component. After the unitary
transformation, we perform a Von Neumann measurement on the
ancillary system. If we get $|P_i\rangle$, $(i=1,2,\cdots, n)$, we
are able to claim that the input state belongs to $S_i$; but if we
get $|P_{n+1}\rangle$, the classification fails. The failure
probability of classifying these states is $\gamma_{ik}$. By
Eq.(10), we have
\begin{eqnarray}
\langle\psi_{ik}|\psi_{jl}\rangle=\sqrt
{(1-\gamma_{ik})(1-\gamma_{jl})}\langle\psi'_{ik}|\psi'_{jl}\rangle\langle P_i|P_j\rangle+\nonumber\\
\sqrt {\gamma_{ik}\gamma_{jl}}\langle\phi_{ik}|\phi_{jl}\rangle.
\end{eqnarray}
For $i\neq  j$,
\begin{equation}
|\langle\psi_{ik}|\psi_{jl}\rangle|=\sqrt {\gamma_{ik}\gamma_{jl}}|\langle\phi_{ik}|\phi_{jl}\rangle|.
\end{equation}
Thus
\begin{equation}
\sqrt {\gamma_{ik}\gamma_{jl}}\geq |\langle\psi_{ik}|\psi_{jl}\rangle|.
\end{equation}
Eq.(13) gives a bound of failure probability of classifying arbitrary two quantum states. One may solve a series
of inequalities to obtain individual failure probabilities. However, we are interested in the average failure
probability. According to Eq.(13), it is easy to obtain
\begin{equation}
\frac {\eta_{ik}\gamma_{ik}}{N-m_i}+\frac {\eta_{jl}\gamma_{jl}}{N-m_j}\geq 2\sqrt {\frac
{\eta_{ik}\eta_{jl}}{(N-m_i)(N-m_j)}}|\langle\psi_{ik}|\psi_{jl}\rangle|.
\end{equation}
So the average failure probability is
\begin{equation}
Q=\sum_{i,k}\eta_{ik}\gamma_{ik}\geq \sum_{i\neq j}\sum_{k,l}\sqrt {\frac
{\eta_{ik}\eta_{jl}}{(N-m_i)(N-m_j)}}|\langle\psi_{ik}|\psi_{jl}\rangle|.
\end{equation}
The average success probability is
\begin{equation}
P=1-Q\leq1-\sum_{i\neq j}\sum_{k,l}\sqrt {\frac
{\eta_{ik}\eta_{jl}}{(N-m_i)(N-m_j)}}|\langle\psi_{ik}|\psi_{jl}\rangle|.
\end{equation}

 This shows that if the subsets are orthogonal, the average success probability will be always 1. The second
 term in Eq.(16) represents the derivation due to the non-orthogonal nature of the states from different
 subsets.

 \section{Conclusion}

 We have considered the problem of general conclusive quantum state classification, the necessary and sufficient
 conditions for the existence of conclusive classification strategies has been presented. Moreover, we have given
 the upper bound for the maximal success probability.

 \acknowledgments This work was supported by Hebei Natural Science Foundation of China under Grant Nos:
A2004000141 and A2005000140, and Key Natural Science Foundation of
Hebei Normal University.

\end{document}